\documentclass[twocolumn,superscriptaddress,amssymb,amsmath,aps,longbibliography,nofootinbib,prx,floatfix]{revtex4-2}
\usepackage{graphicx}
\usepackage{dcolumn}
\usepackage{bm}
\usepackage{float}
\usepackage{comment}
\usepackage{amsmath}
\usepackage{amssymb}
\usepackage{color}
\usepackage{subeqnarray}
\usepackage[utf8]{inputenc}
\usepackage[english]{babel}
\usepackage{mathrsfs}
\usepackage[bb=boondox]{mathalfa}
\usepackage{xcolor} 

\usepackage{amsbsy}
\usepackage{latexsym,epsfig,graphicx}
\usepackage{subfigure}
\usepackage{amsfonts}
\usepackage{amsmath}
\usepackage{xspace}
\usepackage{epstopdf}
\usepackage{array,booktabs}
\usepackage{babel}
\usepackage{multirow}
\usepackage[colorlinks=true, pdfstartview=FitV, linkcolor=red, citecolor=blue, urlcolor=blue]{hyperref}
\usepackage[normalem]{ulem}
\usepackage{cleveref,braket,xcolor}

\pdfoutput=1

\usepackage[utf8]{inputenc}
\usepackage[T1]{fontenc}
\usepackage{mathptmx}
\usepackage{etoolbox}

\makeatother

\begin{document}

\title[AVS Quantum Science]{S-QGPU: Shared Quantum Gate Processing Unit for Distributed Quantum Computing}

\author{Shengwang Du}
\email{dusw@purdue.edu}
\affiliation{Elmore Family School of Electrical and Computer Engineering, Purdue University, West Lafayette, Indiana 47907, USA}
\affiliation{Department of Physics and Astronomy, Purdue University, West Lafayette, Indiana 47907, USA}
\affiliation{Department of Physics, The University of Texas at Dallas, Richardson, Texas 75080, USA}

\author{Yufei Ding}
\email{yufeiding@ucsd.edu}
\affiliation{Department of Computer Science and Engineering, University of California San Diego, La Jolla, California 92093, USA}

\author{Chunming Qiao}
\email{qiao@buffalo.edu}
\affiliation{Department of Computer Science and Engineering, University at Buffalo, The State University of New York, Buffalo, New York 14260, USA}

\date{\today}

\begin{abstract}
We propose a distributed quantum computing (DQC) architecture in which individual small-sized quantum computers are connected to a shared quantum gate processing unit (S-QGPU). The S-QGPU comprises a collection of hybrid two-qubit gate modules for remote gate operations. In contrast to conventional DQC systems, where each quantum computer is equipped with dedicated communication qubits, S-QGPU effectively pools the resources (\textit{e.g.}, the communication qubits) together for remote gate operations, and thus significantly reduces the cost of not only the local quantum computers but also the overall distributed system. Our preliminary analysis and simulation show that  S-QGPU's shared resources for remote gate operations enable efficient resource utilization. When not all computing qubits (also called data qubits) in the system require simultaneous remote gate operations, S-QGPU-based DQC architecture demands fewer communication qubits, further decreasing the overall cost. Alternatively, with the same number of communication qubits, it can support a larger number of simultaneous remote gate operations more efficiently, especially when these operations occur in a burst mode.
\end{abstract}

\maketitle


\section{Introduction}\label{sec:Introduction}

\begin{figure*}[t]
\centering
\includegraphics[width=1.0 \textwidth]{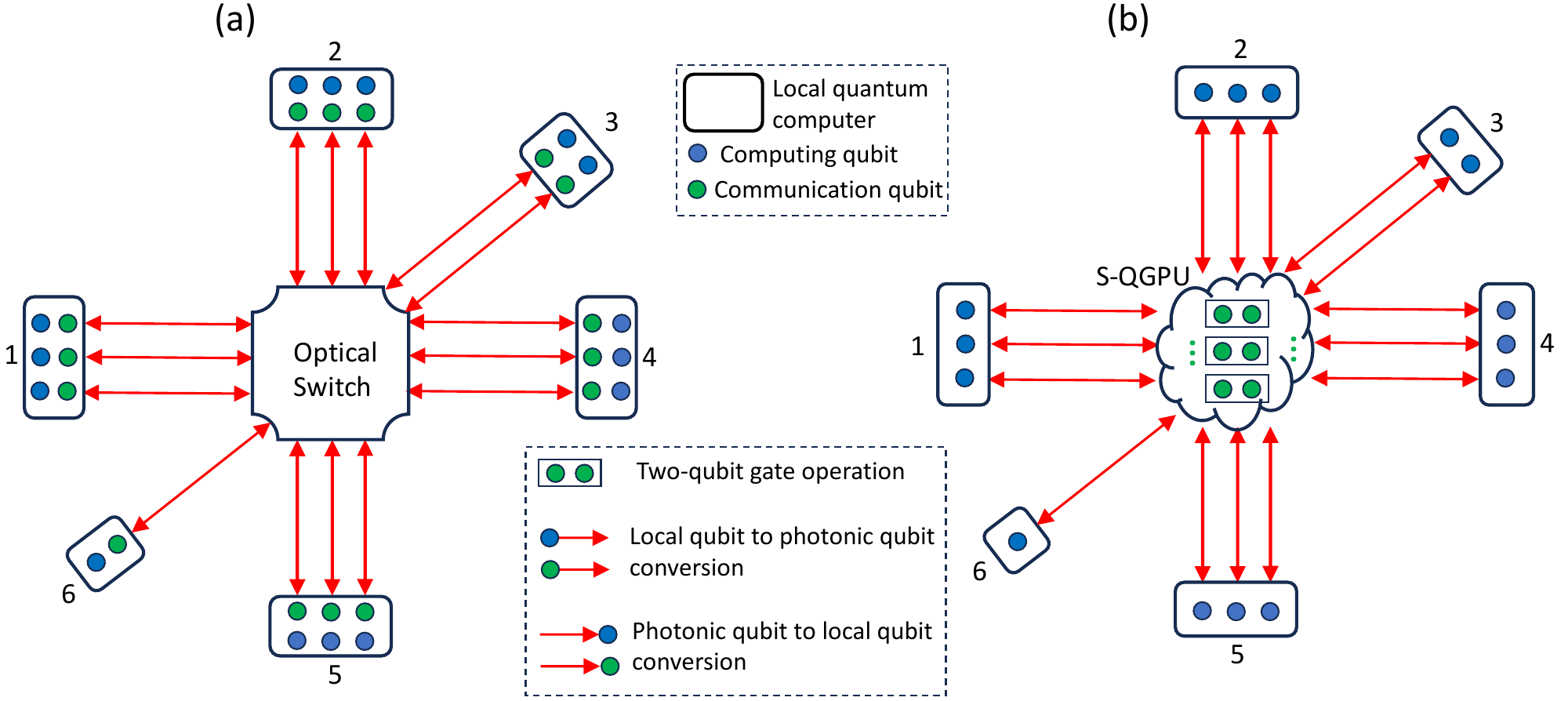}
\caption{Two distributed quantum computing architectures. (a) Conventional entanglement-communication-based architecture where each node has both computing qubits and communication qubits, and the nodes are interconnected via a quantum network (shown as a switch and quantum links for simplification). (b) S-QGPU-based architecture, where all communication qubits, paired into a collection of two-qubit gate units, are shared by the nodes which have only computing qubits.}
\label{fig:SGD}
\end{figure*}

Quantum computing represents a paradigm shift in computational power, promising to solve problems that are currently intractable for classical computers. Theoretically, an ideal quantum computer with 300 qubits possesses a staggering $2^{300}$-dimensional Hilbert space, rendering classical computing techniques inaccessible, even if every atom in the observable universe were to be converted into classical bits. Such a fully-connected $N$-qubit gate-based universal quantum computer requires that a unitary controlled gate operation can be directly performed between any two arbitrary qubits. However, in the real world, the connectivity of a local quantum computer is always limited by physical constraints. Therefore, the number of qubits in a quantum computer is not a ``precise'' measure of its overall computing power \cite{IBMQuantumVolume}. For example, in a superconductor quantum computer, its real computing power is always much lower than what appears to be based on the number of qubits, because a superconductor qubit can only interact with its nearby qubits (nearest-neighbor coupling), and hence, completion of a gate operation between two distant qubits must go through many sequential intermediate gates \cite{npjQI-Matthias}. For this reason, although IBM has released a 1121-qubit processor called Condor in 2023 \cite{Gambetta2023}, they are all still far away from being sufficient to support practically useful applications beyond scientific interests. Trapped-ion \cite{IonQC-Sage} and single neutral-atom-based \cite{PhysicsToday-Saffman, PhysRevLett.104.010503} quantum computers have similar challenges.

Due to many physical constraints and technological limitations, it is difficult, if not impossible, to build a monolithic fully-connected quantum computer with a very large number ($N$) of qubits, and let alone to guarantee a direct controlled gate operation can be performed between any two arbitrary qubits. Extending from $N$ to $N+1$ in such a quantum computer is more than just physically adding one more qubit. Therefore, a conventional wisdom is that the cost of such a fully-connected local quantum computer increases \emph{exponentially} as the number of qubits increases. Consequently, there is a growing interest in exploring distributed quantum computing (DQC) systems that can interconnect many small-size, cost-effective local quantum computers or processors \cite{SciRep2021115172, 8910635, 7562346, iet-qtc.2020.0002, PhysRevA.104.052404, zhao2021quantum, Cat-Comm, PhysRevA.62.052317}. 

Nevertheless, the central challenge in constructing an effective DQC system lies in facilitating remote gate operations that involve computing qubits from different nodes. To accomplish this, several components are typically required in addition to a quantum network comprising quantum links, switches, and repeaters to support inter-node quantum communication \cite{KimbleQN}. These components include (1) transducers for efficient conversion between local qubits and flying photonic qubits, and (2) auxiliary qubits, referred to as \emph{communication qubits} hereafter, used primarily to temporarily store the quantum state information for remote gate operations. It's worth noting that when two nodes are connected, each equipped with $N$ computing qubits, via classical means, the dimension of the combined Hilbert space is, at most, $2 \times 2^N = 2^{(N+1)}$. This is significantly smaller than the $2^{2N}$ dimension achievable by an ideal DQC system with these two nodes.

In most conventional DQC architectures, each local quantum computer is equipped with additional communication qubits dedicated to establishing remote entanglement links \cite{Cat-Comm, PhysRevA.62.052317}. The presence of these communication qubits substantially escalates the cost of individual local quantum computer nodes. Furthermore, in these schemes, remote gate operations based on entanglement are post-selected via measurement, rendering them inherently non-deterministic. In this study, we introduce a novel DQC architecture based on \emph{Shared Quantum Gate Processing Unit} (S-QGPU), built upon a recent breakthrough concept known as the hybrid two-qubit gate module \cite{quteDu202300007}. This module not only performs transduction between local qubits and photonic qubits but also facilitates controlled gate operations. In the S-QGPU-based architecture, nodes are interconnected to a centralized shared collection of these hybrid two-qubit gate modules, where all remote gate operations are executed. In contrast to the conventional architectures, the implementation of S-QGPU leads to a significant reduction in the costs associated with local quantum computing nodes. Moreover, the remote gate operations conducted within the S-QGPU framework are deterministic, eliminating the inherent non-determinism found in traditional DQC architectures.

This article is structured as follows. In Sec.~\ref{sec:SQGPU}, we describe the DQC architecture based on S-QGPU. In Sec.~\ref{sec:special-case}, we analyze the cost of S-QGPU-based scheme with full pairing capacity, as a comparison with that of the corresponding conventional scheme as well as a monolithic quantum computer. In Sec.~\ref{sec:general-analysis}, we extend our cost analysis to a more general case where not all the computing qubits are remotely paired. In Sec.~\ref{sec:burst-performance}, we evaluate and compare the performances of S-QGPU-based and conventional architectures when the inter-node communication pattern is burst. Finally, we conclude in Sec.~\ref{sec:conclusion}.

\section{Architecture Models}\label{sec:SQGPU}

To describe our proposed S-QGPU-based architecture for DQC, we first look at the conventional entanglement-communication-based scheme, as illustrated in Fig.~\ref{fig:SGD}(a). In this setup, each local node is equipped with not only computing qubits, but also communication qubits for supporting remote gate operations. To perform a remote (2-qubit) controlled gate operation involving the first computing qubit, denoted by $p$ on node 1, and the second computing qubit, denoted by $q$, on node 2, we first entangle a communication qubit, $p'$, on node 1 with a communication qubit, $q'$, on node 2. Then, the remote gate operation between the two computing qubits $p$ and $q$ can be accomplished with local operations involving $p$ and $p'$ on nodes 1, followed by transmission of the Bell state measurement of the resulting $p'$ from node 1 to node 2 using a classical channel, and finally, another local operation involving $q$ and $q'$ on node 2 based on the received measurement result. In this architecture, the nodes need to be interconnected with each other via a quantum network consisting of links and switches to route (or distribute) entanglement. In addition, establishing entanglement between two nodes typically involves transduction at both the sources that generate entangled photon pairs, and each (destination) node that stores one of the entangled photons using a communication qubit. This entanglement-communication-based DQC architecture can be implemented with either the widely recognized Cat-entangler scheme or the teleportation scheme \cite{PhysRevA.62.052317, Cat-Comm}, which are inherently non-deterministic as their remote gate operations are post-selected based on measurement. 

Our proposed S-QGPU-based architecture is depicted in Fig.~\ref{fig:SGD}(b). In this new architecture, each node has computing qubits but no communication qubits are needed to support remote gate operations. Instead, it uses a collection of hybrid two-qubit gates \cite{quteDu202300007} called S-QGPU as shared resources to support remote gate operations. All the nodes are connected to the centralized S-QGPU, and transduction is performed at all the nodes and the S-QGPU. More specifically, to perform a remote gate operation involving computing qubit $p$ on node 1, and computing qubit $q$ on node 2, we first convert $p$ and $q$ into flying photonic qubits (a process often known as quantum transduction, which in this case is based on atom-photon interactions), which are subsequently transmitted to any one of many hybrid two-qubit gate modules, available at S-QGPU for processing. Upon completion of the remote gate operation, the module (in S-QGPU) then returns the resulting two qubits (also as photons) to nodes 1 and 2, respectively. In addition, the module will be reset, so that it can be used to perform another remote gate operation involving any two computing qubits on any two nodes in the future. In S-QGPU-based architecture, the remote gate operations are deterministic.

Note that S-QGPU-based DQC architecture requires (1) efficient conversion (transduction) between local computing qubits and flying photonic qubits; and (2) each two-qubit gate module capable of performing a complete set of $4\times 4$ unitary operations.  While we are not limiting system implementation to any particular platform, the atom-photon hybrid modules proposed by Oh \textit{et al.} \cite{quteDu202300007} could meet these requirements. In addition, from a technological point of view, S-QGPU requires more advanced implementation than the conventional entanglement-communication-based architecture, and in particular, the qubits in the hybrid modules are more functional than typical communications qubits. Nevertheless, for ease of presentation, we still treat the qubits in the hybrid modules to be the same as the communication qubits in the following discussions. 

In this work, we will compare the proposed S-QGPU-based DQC with the conventional entanglement-communication-based architecture in terms of cost and performance. We show that for an ideal system with $M$ nodes, each having $N$ computing qubits, and total having $MN$ communication qubits to accommodate full pairing capacity, S-QGPU-based system results in a lower cost when $M$ and $N$ are sufficiently large. This is because the dependence of the cost of a local quantum computer on its number of qubits is nonlinear, while the cost of S-QGPU is linear on its number of modules. The cost of transducers, quantum communication channels, and optical switches, have linear or polynomial cost functions of $M$ and $N$.

In addition, we also show that when not all $MN$ computing qubits need to be engaged in remote gate operations at the same time, S-QGPU-based architecture can realize a more significant reduction in the number of communication qubits than what is possible with the conventional entanglement-communication-based architecture, and accordingly, additional cost-savings on top of an already lower cost. Moreover, in terms of performance comparison, we show that when the number of communication qubits is limited (and the same) in both architectures, S-QGPU-based scheme is more efficient when remote gate operations are unevenly distributed at either the computing qubit level or the node level, or in other words, when the pattern for inter-node communication (supporting remote gate operations) is burst. Such a more significant reduction in the number of communication qubits, and accordingly additional reduction in the cost-savings, as well as performance improvement under a bursty communication pattern, all come as a well-recognized benefit from sharing of the resources, which, in our context, refers to the communication qubits (and the hybrid two-qubit gate modules) supporting remote gate operations. We provide a detailed analysis in the following sections.

\section{Cost Comparison: A special case with full pairing}\label{sec:special-case}

In this section, we consider a distributed system with $M$ nodes, each having $N$ computing qubits. To simplify the presentation, we assume that $MN$ is an even number and consider the case where all $MN$ computing qubits can be fully paired for remote gate operations at the same time, which requires $MN/2$ hybrid two-qubit gate modules in S-QGPU. Based on the conventional wisdom mentioned earlier, the cost of a fully-connected $N$-qubit local quantum computer is estimated as $\epsilon (a^N-1)$, where $\epsilon > 0$ and $a > 1$ are cost parameters (coefficients), whose exact values may change as the technology advances. With the S-QGPU being an assembly of $MN/2$ hybrid modules, each being two-qubit gates, its estimated cost is $(MN/2)\epsilon(a^2-1)$ (this is because the cost of each two-qubit gate can be considered as being equivalent to that of a two-qubit node).  To ensure full pairing interconnectivity of the total $MN$ computing qubits, $MN$ quantum communication channels are required between the local nodes and the S-QGPU, so is an $MN \rightarrow MN$ optical switch. The quantum communication channel cost, including conversion between local qubits and photonic qubits, is approximated as $MNb$, with $b > 0$ indicating the cost for each quantum channel. The cost of an $MN \rightarrow MN$ optical switch, which is purely classical, is estimated as $(MN)^2d$ where $d > 0$ is the cost per switching path. The overall cost, denoted by $C_S$, can be formulated as follows:
\begin{equation}
\begin{aligned}
C_{S}(M,N) =& M\epsilon (a^N-1) + \frac{1}{2} MN\epsilon(a^2-1) \\
+&MNb+(MN)^2d.
\end{aligned}
\label{eq:Cost1}
\end{equation}
For the conventional entanglement-communication-based architecture shown in Fig.~\ref{fig:SGD}(a), it requires $N$ communication qubits per node. Because each node has $2N$ qubits to accommodate both computation and communication, the cost of a system, denoted by $C_E$, is given by:
\begin{equation}
\begin{aligned}
C_{E}(M,N) =M\epsilon (a^{2N}-1) + MNb+ (MN)^2d.
\end{aligned}
\label{eq:Cost2}
\end{equation}
For reference, the cost of building a monolithic fully-connected local quantum computer with $MN$ qubits, denoted by $C_{0}$, is:
\begin{equation}
\begin{aligned}
C_{0}(M,N) =\epsilon (a^{MN}-1).
\end{aligned}
\label{eq:Cost3}
\end{equation}

From the above equations, it is evident that despite the fact that the overall cost of any approach is an exponential function of $N$, both S-QGPU-based and conventional entanglement-communication-based systems have a cost that is polynomial to the number of nodes, $M$, whereas the monolithic computer has a cost that is exponential to $M$ as well. Accordingly, when building a system with a large number of computing qubits (and thus $M$ could be large), both distributed approaches will have a lower cost than the monolithic approach. Among the two distributed architectures, S-QGPU-based has a smaller exponent than the conventional architecture, and accordingly, a lower cost when $N$ is large enough.

To quantitatively analyze and compare the costs, we take the following estimations. Assuming that a two-qubit quantum computer costs \$5,000, and a 50-qubit fully-connected quantum computer costs \$4,000,000, we have $\epsilon=\$21,476$, and $a=1.11032$. For illustration purposes, we (somewhat arbitrarily) assume $b=\$10,000$ per quantum communication channel, and $d=\$100$ per optical switching path.

\begin{figure}[t]
\centering
\includegraphics[width=1.0 \linewidth]{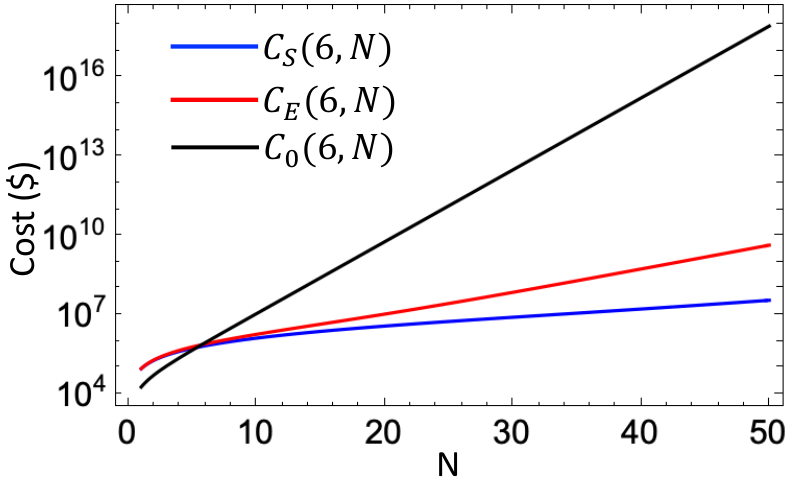}
\caption{Costs $C_S(6,N)$, $C_E(6,N)$, and $C_0(6,N)$ with fixed node number $M=6$ as functions of $N$, the number of computing qubits at each node. Parameters: $\epsilon=$\$21,476, $a=$1.11032, $b=$\$10,000, and $d=$\$100.}
\label{fig:CostCase2}
\end{figure}

\begin{figure}[t]
\centering
\includegraphics[width=0.95 \linewidth]{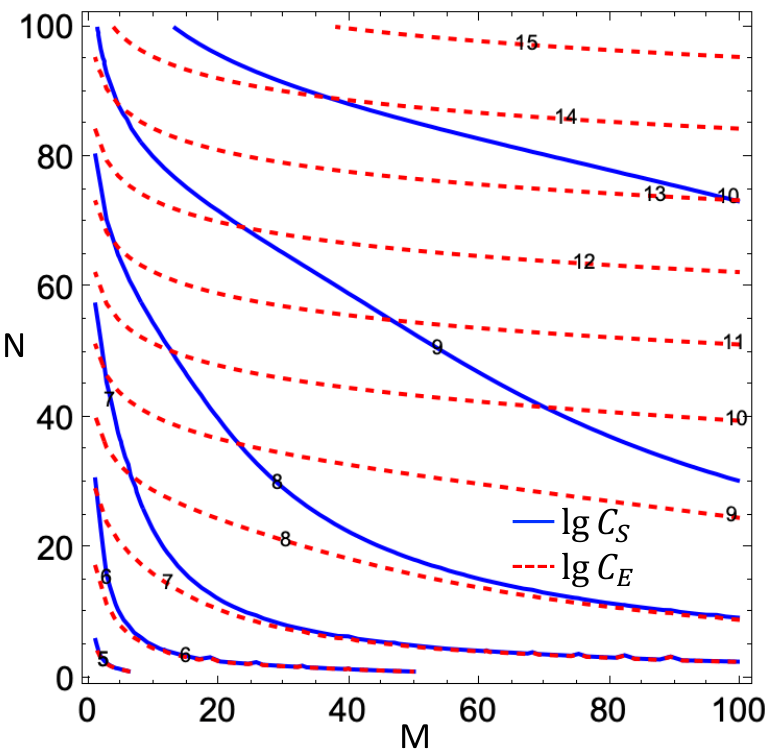}
\caption{Contour plots of $\mathrm{lg}[C_S(M,N)]$ and $\mathrm{lg}[C_E(M,N)]$. The solid (blue) and dashed (red) lines are equal-cost curves for $\mathrm{lg}C_S$ and $\mathrm{lg}C_E$, respectively. Parameters: $\epsilon=$\$21,476, $a=$1.11032, $b=$\$10,000, and $d=$\$100.}
\label{fig:ContourPlots}
\end{figure}

Let us first set $M=6$. As depicted in Fig.~\ref{fig:CostCase2}, substantial cost disparities emerge among the three schemes: $C_S(6,N) < C_E(6,N) < C_0(6,N)$. In this case, the turning point for $C_S\simeq C_L \simeq C_0$ occurs at $N=5$. As $N>5$, the cost of S-QGPU-based scheme starts to reduce cost significantly as compared to the conventional scheme. It is possible to achieve 300-qubit distributed quantum computing at a cost of $\$37$M. 

The costs are jointly determined by both $M$ and $N$, as shown in the contour plots in Fig.~\ref{fig:ContourPlots}, where a number associated with each (log-scale) cost curve represents the exponent associated with the overall system cost. As can be seen, with a small $N$, the cost of S-QGPU-based scheme is nearly identical to that of the corresponding  conventional scheme. As $N$ increases, the cost of the nodes is significantly reduced in the S-QGPU-based architecture, and thus it becomes more cost-effective. At a large $N$, the exponential terms dominate in Eqs. \eqref{eq:Cost1}-\eqref{eq:Cost3}, leading to the approximate relations:
\begin{equation}
\frac{C_S(M,N)}{C_0(M,N)} \simeq M a^{N-MN} \simeq M a^{-MN},
\label{eq:Cost13}
\end{equation}
and
\begin{equation}
\frac{C_S(M,N)}{C_E(M,N)} \simeq a^{N-2N} = a^{-N}.
\label{eq:Cost14}
\end{equation}    

\section{Cost Comparison: A General Case with partial pairing}\label{sec:general-analysis}

So far, we have examined a special case where the DQC system is designed with full pairing capacity such that all the $MN$ computing qubits can be involved in some remote gate operations at the same time. In such a case, the total number of communication qubits needed, denoted by $Q$, is $MN$, and the maximum number of remote gate operations that can be supported concurrently, denoted by $R$, is $MN/2$, in both architectures. Essentially, at each of the $M$ nodes, all its $N$ computing qubits are allowed to be involved in certain remote gate operations at the same time. In this section, we examine a more general case with partial pairing, where the system is designed to support a class of applications that require that (at most) $x \le N$ computing qubits from each of $y \le M$ nodes be involved in remote gate operations at any given time. It is clear that in the special full pairing case studied earlier, we have $x = N$ and $y = M$. Note that when $x < N$ and/or $y < M$, not all $MN$ computing qubits engage in remote gate operations concurrently at the same time. In S-QGPU-based architecture, we just need $R = xy/2\le MN/2$ two-qubit gate units with a total $2R=xy$ equivalent shared communication qubits. In the conventional entanglement-communication-based architecture, however, as we do not know which $y$ (of $M$) nodes are involved in simultaneous remote gate operations, we have to assume all $M$ nodes possibly engage and thus assign redundant communication qubits.

We note that not all applications require $MN$ computing qubits be fully paired in remote gate operations at the same time. Accordingly, if none of the applications to be run on the system would require more than $R$ pairs of remote gate operations simultaneously, then one may consider the system's effective capacity to be large enough to create an illusion that its computing capacity is on the order of $2^{MN}$.

In the following discussion, let $Q_E$ and $Q_S$ be the total number of communication qubits needed by the system with entanglement-communication-based and S-QGPU-based architectures respectively. As to be shown next,  when $x < N$ and/or $y < M$, we have $Q_E > Q_S$, although in the special case (where $x = N$ and $y = M$), we have $Q_E = Q_S = MN$. This means that in general, the cost saving achieved by S-QGPU-based architecture over the conventional entanglement-communication-based architecture can be even more significant than in the special full pairing case. 

Below, we discuss the general partial pairing case where $Q_E, Q_S \ge xy = 2R$ in more detail. Assuming as before that the nodes are homogeneous (i.e., having the same configuration in terms of the number of computing and communication qubits), we examine two subcases: 
\begin{enumerate}
    \item \textbf{Subcase E}: Evenly-distributed remote gate operations. In this subcase, given at most $R=  xy/2$ simultaneous remote gate operations at any given time, there are up to $y \le M$ nodes that are engaged in concurrent remote gate operations. In addition, among these $y$ nodes, the same number of computing qubits at \emph{each} node, up to $x$, is involved in remote gate operation at the same time as other computing qubits at other $y-1$ nodes. In other words, the computing qubits involved in remote gate operations are evenly distributed over the $y$ nodes.
    \item \textbf{Subcase U}: Unevenly distributed remote gate operations. In this subcase, for a given $R \leq N$, it is possible that at only one node, all $R$ computing qubits are engaged in remote gate operations, whereas the other $R$ computing qubits are arbitrarily distributed over the remaining $y-1$ nodes. If $R > N$, then it is possible that at one or a few nodes, all $N$ computing qubits at each node are engaged in remote gate operations, whereas at some other nodes, fewer computing qubits are engaged in remote gate operations concurrently.  In other words, the computing qubits involved in remote gate operations are unevenly distributed over the $y$ nodes.
\end{enumerate}

\begin{figure}[t]
\centering
\includegraphics[width=1 \linewidth]{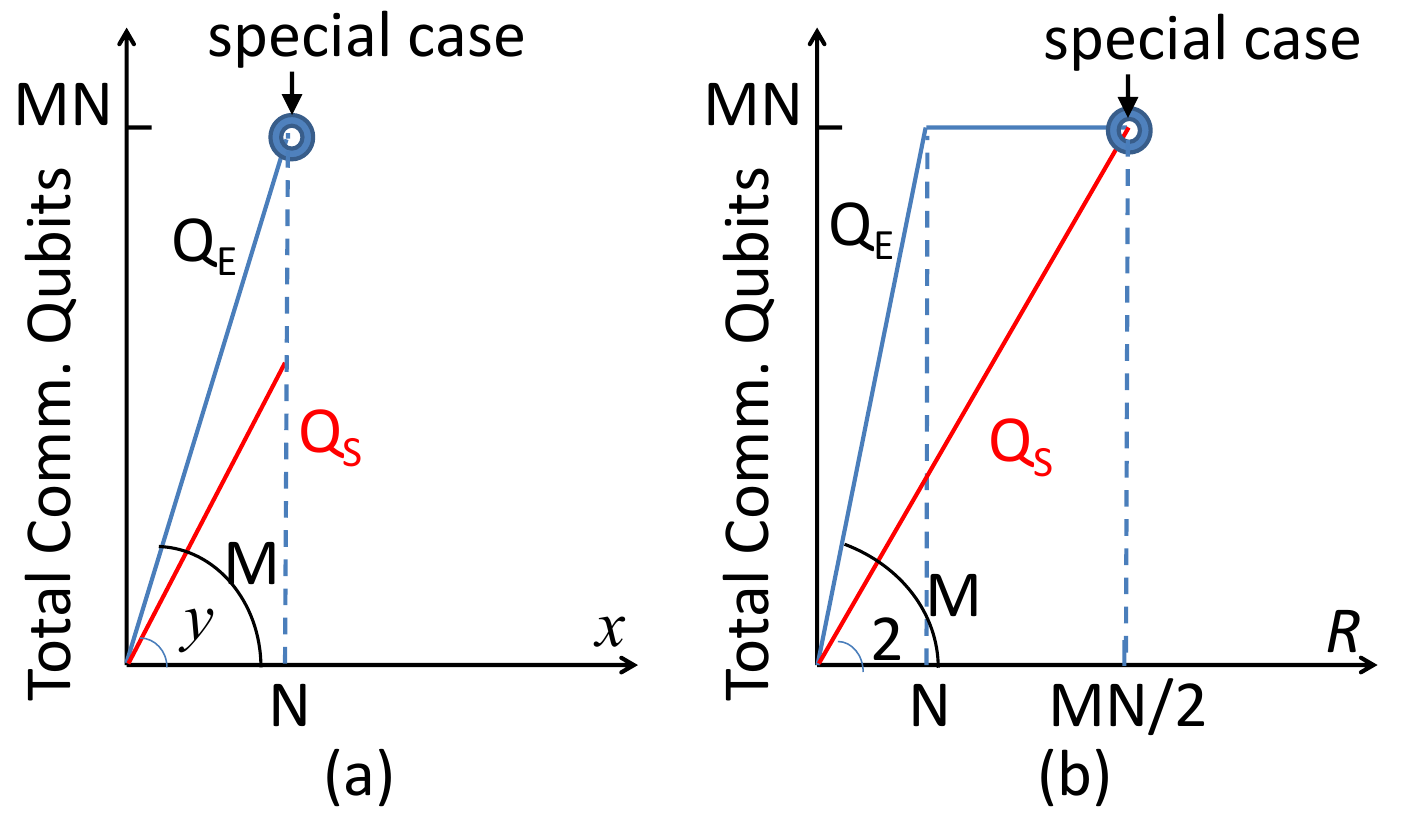}
\caption{Comparing the number of communications qubits needed in two architectures in the general case. (a) Subcase E, $Q_E$ vs $Q_S$ as functions of $x$. (b) Subcase U, $Q_E$ vs $Q_S$ as functions of $R$}.
\label{fig:subcaseEvsU}
\end{figure}

In Subcase E, the conventional entanglement-communication-based architecture would need to dedicate $x$ communication qubits at each node, resulting in a total of $Q_{E} = Mx$ communication qubits as we do not know which $y$ nodes would engage, whereas in the proposed S-QGPU-based architecture, the total number of communication qubits needed is only $Q_{s} = 2R = xy$. When $y < M$, it is clear that $Q_S < Q_E$. As shown in Fig.~\ref{fig:subcaseEvsU}(a), $Q_S$ and $Q_E$ follow different slopes as functions of $x$.

In Subcase U, when $R \le N$, the conventional entanglement-communication-based architecture would need to dedicate $R$ communication qubits at each node, resulting in a total of $Q_{E} = MR$ communication qubits, whereas in the proposed S-QGPU-based architecture, the total number of communication qubits needed is still  $Q_{S} = 2R$ only. When $R > N$, the number of communication qubits needed at each node in entanglement-communication-based architecture plateaus at $N$, resulting in a total of $Q_{E} = MN$ communication qubits, whereas in the proposed S-QGPU-based architecture, the total number of communication qubits needed remains as $Q_{S} = 2R$ only. The comparative results are illustrated in Fig.~\ref{fig:subcaseEvsU}(b).

We can generalize the cost function in Eq.~(\ref{eq:Cost1}) as follows (for both subcases E and U):
\begin{equation}
\begin{aligned}
C^{EU}_{S}(M,N) =& \epsilon M(a^N-1) + \epsilon R(a^2-1) \\
&+ bMN+d(MN \cdot 2R).
\end{aligned}
\label{eq:ReducedCost1}
\end{equation}
Similarly, in Subcase E, we can generalize  Eq.~(\ref{eq:Cost2}) as follows:
\begin{equation}
\begin{aligned}
C^E_{E}(M,N) =\epsilon M(a^{N+x}-1) + bMN+ d (MN \cdot Mx).
\end{aligned}
\label{eq:ReducedCost2E}
\end{equation}
and in Subcase U  we can generalize  Eq.~(\ref{eq:Cost2}) as follows when $R \le N$ (note that when $R > N$, there is no change to  Eq.~(\ref{eq:Cost2}).
\begin{equation}
\begin{aligned}
C^U_{E}(M,N) =\epsilon M(a^{N+R}-1) + b MN+ d (MN \cdot MR).
\end{aligned}
\label{eq:ReducedCost2U}
\end{equation}

\section{Performance Comparison with Burst Remote Gate Operations}\label{sec:burst-performance}

In addition to the notable size and cost scalability of the proposed S-QGPU-based architecture, this section delves into showcasing its heightened performance scalability across various ranges of bursty remote gate operations. Bursty inter-processor communication~\cite{farrington2010helios, xu2001techniques,yoo2006optical} is prevalent in classical data centers~\cite{shan2018micro,benson2010understanding, alizadeh2010data} and distributed computing systems~\cite{alakeel2010guide,drougas2009accommodating,abramson1970aloha}, and we anticipate a similar phenomenon in DQC.  In particular, we consider both qubit-level and node-level burstiness. With the former, each computing qubit will engage in remote gate operations at any given time with a certain probability. With the latter, each node will engage in remote gate operations with other nodes at any given time with a certain probability. Here, when a node engages in remote gate operations, all of its computing qubits are engaged in remote gate operations.
From the perspective of the quantum network that interconnects different nodes with each other in the conventional entanglement-communication-based architecture, or with the S-QGPU in the proposed architecture,
a higher probability at either the qubit-level or node-level represents a heavier communication load and higher degree of burstiness, resulting in a higher demand for the use of communication qubits to support remote gate operations.

With a limited number of communications qubits, it is likely that not all of the desired remote gate operations can be performed at once. Accordingly, it may take some computing qubits multiple time steps to complete its remote gate operations, resulting in a longer communication latency. In this section,  we conduct simulations to evaluate the respective performance of S-QGPU-based and the conventional entanglement-communication-based architecture by assuming that they have the same number of nodes ($M$) which is a parameter in the simulation, with the same (and fixed) number of computing qubits per node ($N=50$), and the same total number of communication qubits $Q_E = Q_S = 10M$. 

\begin{figure}[t]
\centering
\includegraphics[width=0.98 \linewidth]{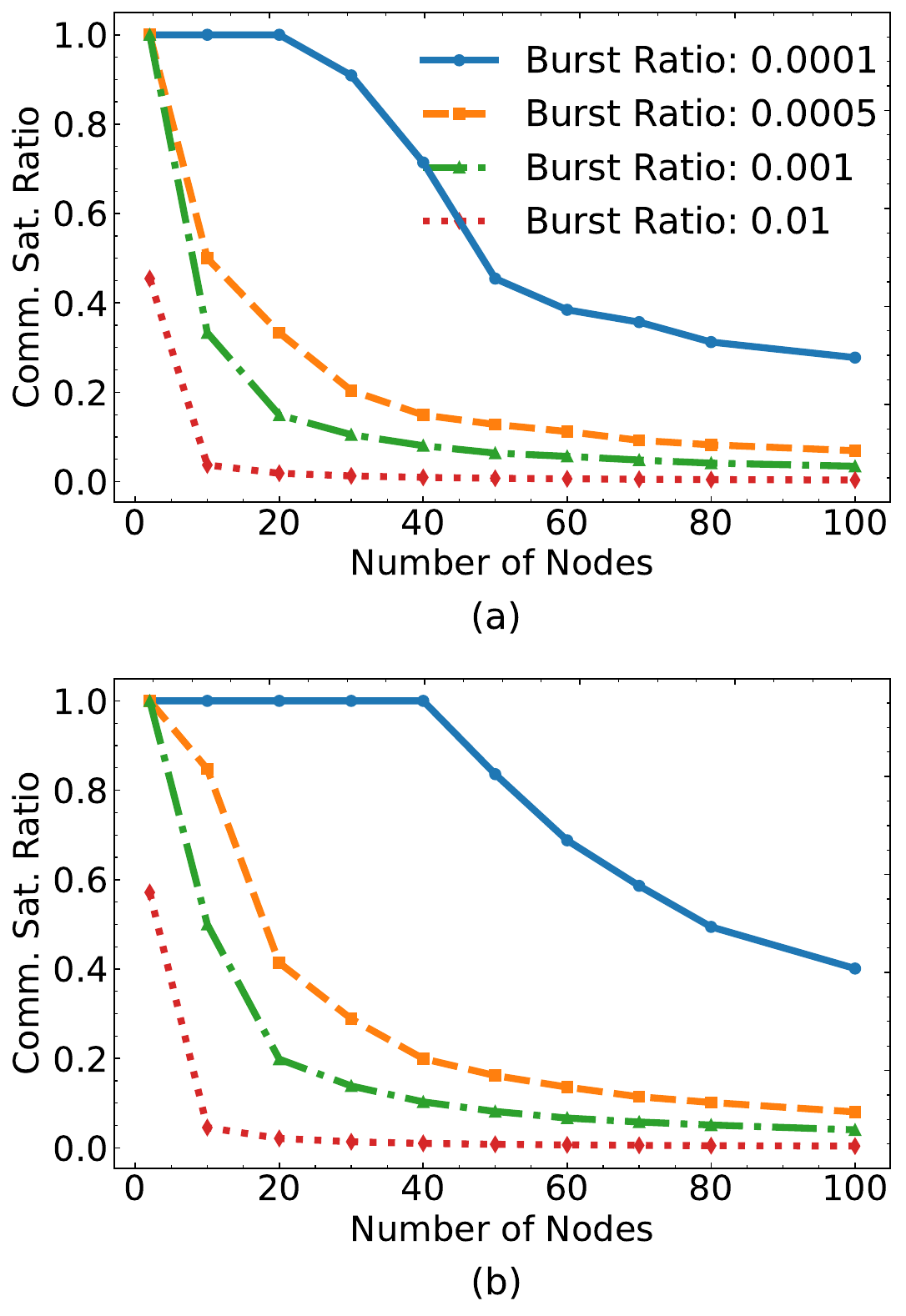}
\caption{Comparison of communication satisfaction ratio (CSR) between dedicated and shared qubits under qubit-level burst communication when scaling up with the number of nodes. (a) The conventional entanglement-communication-based architecture with dedicated communication qubits. (b) The S-QGPU-based architecture with shared communication qubits. Legend for (b) mirrors (a).}
\label{fig:qubit-level_comparision}
\end{figure}

\begin{figure}[t]
\centering
\includegraphics[width=0.98 \linewidth]{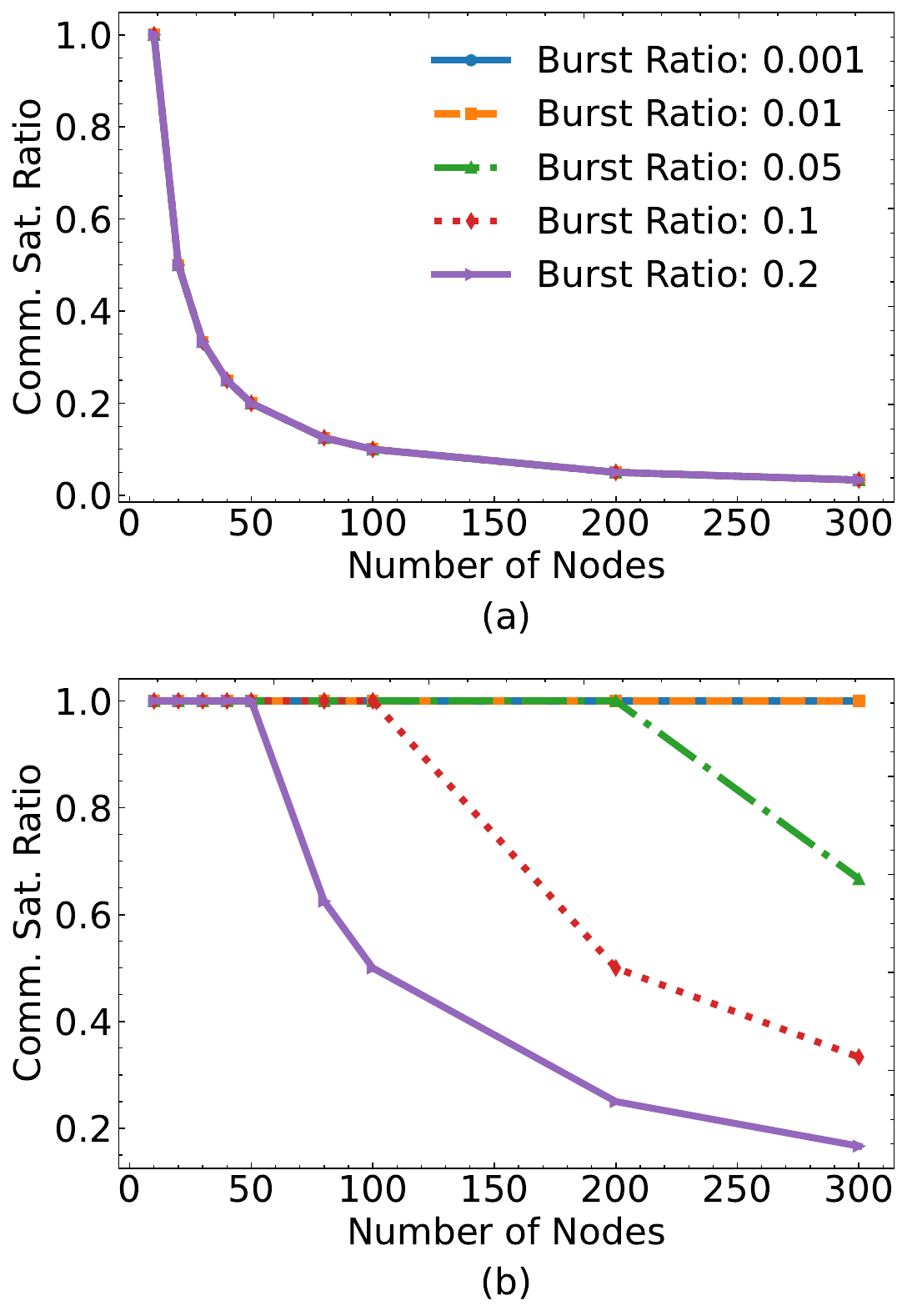}
\caption{Comparison of communication satisfaction ratio between dedicated and shared qubits under node-level burst communication when scaling up with the number of nodes. (a) The conventional entanglement-communication-based architecture with dedicated communication qubits. (b) The S-QGPU-based architecture with shared communication qubits. Legend for (b) mirrors (a).}
\label{fig:node-level_comparision}
\end{figure}

Our assessment of performance under bursty communication patterns involves determining the latency incurred for all computing qubits to complete their requests for remote gate operations that are generated at a given time, providing profound insights into the operational prowess and efficiency of an architecture under dynamic workloads (in terms of remote gate operations).
Specifically, let \( L_r \) be the average latency under realistic settings (with a limited $Q_E = Q_S = 10M$) and \( L_i \) be the latency in the ideal setting (with an unlimited number of communication qubits), we define the Communication Satisfaction Ratio (CSR) to be:
\begin{equation}
\begin{aligned}
\text{CSR} = \frac{L_r}{L_i}.
\end{aligned}
\label{eq:CSR}
\end{equation}

Figure~\ref{fig:qubit-level_comparision} gives the comparison of communication efficiency between entanglement-communication based architecture (where each of the $M$ nodes has 10 dedicated communication qubits), and the proposed S-QGPU-based architecture (where $10M$ communication qubits are shared among the $M$ nodes) under qubit-level burst communications. As mentioned earlier, here, the burst ratio encapsulates the likelihood of a particular computing qubit actively participating in remote gate operation during each and every discrete time step. Specifically, a local computing qubit has a probability of \(x\%\) to initiate a remote gate operation with any qubits on other nodes. In the conventional entanglement-communication based architecture, such a remote gate operation leverages a communication qubit from both the source node and the target node (for creating the entanglement). 
In contrast, in the proposed S-QGPU shared architecture, such an operation requires a qubit pair from the shared pool. 

In the simulations, we systematically vary $M$ while keeping both $N$ and $Q$. Two subfigures illustrate the results for different communication settings: Fig.~\ref{fig:qubit-level_comparision} (a) the conventional entanglement-communication based architecture with dedicated communication qubits on each node, and Fig.~\ref{fig:qubit-level_comparision} (b) our S-QGPU-based architecture with a pool of two-qubit gate modules shared among all the nodes. Each curve represents a specific burst ratio at the qubit level and shows how the Communication Satisfaction Ratio varies with the number of nodes. It is evident that our S-QGPU-based architecture consistently outperforms its counterpart. Regardless of the number of nodes considered, the S-QGPU-based architecture consistently achieves a higher CSR. Moreover, our architecture enables scaling to a greater number of nodes while maintaining a desired CSR. This performance superiority is attributed to the intrinsic efficiency derived from sharing communication qubits as resources in our S-QGPU-based architecture, manifested further in the presence of bursty remote gate operation patterns with limited resources. 

Figure~\ref{fig:node-level_comparision} further gives the comparison of communication efficiency between the two architectures under node-level burst communication. 
As mentioned earlier, the node-level burst ratio signifies the probability of an entire node engaging in remote communication with all other nodes during discrete time steps. This abstraction captures the collective activity of a node as a whole, encompassing all the individual qubits within the node. Node-level burstiness amplifies the importance of the core advantage of sharing communication qubits across nodes, offered by the S-QGPU-based architecture. As shown in Fig.~\ref{fig:node-level_comparision} (a), the conventional approach reliant on dedicated communication qubits at each node completely fails to adapt to dynamic communication requirements across different nodes, resulting in a low CSR when $M$ is large, even with a relatively low burst ratio, indicating consistent bottlenecks created by nodes with the highest communication demands. On the other hand, as shown in Fig.~\ref{fig:node-level_comparision} (b), the inherently flexible resource allocation in our S-QGPU-based architecture empowers it to deliver enhanced scalability to serve a larger number of nodes with a reasonably low burst ratio.

\section{Conclusion}\label{sec:conclusion}
In summary, we have introduced a new DQC architecture, which leverages multiple small quantum computer nodes to achieve a high quantum computing power by interconnecting them to a \emph{shared} quantum gate processing unit (S-QGPU) that performs all remote gate operations.  Unlike conventional DQC architectures based on entanglement communication, wherein remote gate operations are accomplished via teleportation or cat-entanglers \cite{Cat-Comm,PhysRevA.62.052317}, the proposed S-QGPU approach for remote gate operations is deterministic and does not depend on any measurement-based post selection. 

The proposed S-QGPU-based architecture offers a more scalable solution with improved cost-effectiveness and operational efficiency, when the number of fully-connected qubits on each quantum computer is limited by technology. Through extensive cost analysis, we have shown that S-QGPU-based DQC architecture significantly reduces both cost and resource requirements in both full and partial pairing scenarios compared to the conventional entanglement-communication-based architecture. In addition, our simulation-based performance evaluation has further revealed the superior communication efficiency of S-QGPU-based architecture in handling bursty requests for remote gate operations both at the qubit-level and the node-level.  Overall, this study has demonstrated a tremendous potential to both scale up and scale out  quantum computing power significantly using the S-QGPU-based architecture, though much further research still needs to be done. 

Finally, we also note that although a hybrid two-qubit gate module with efficient transduction interfaces between local qubits and flying photonic qubits \cite{quteDu202300007} is currently being considered a key element of the envisioned S-QGPU, the proposed concept of sharing resources (such as these modules) itself to enable a novel DQC architecture applies equally well to other qubit and transduction technologies. For example, the S-QGPU architecture can also be implemented in superconductor quantum computing platform with microwave links to overcome the constraints on qubit connectivity \cite{pnas.1618020114, npjQinf6672020}.

\begin{acknowledgments}
S.D. acknowledges support from AFOSR (FA9550-22-1-0043), DOE (DE-SC0022069) and NSF (2114076, 2228725). Y.D. acknowledges support from NSF (2048144), Cisco Research, and Robert N. Noyce Trust.
\end{acknowledgments}

\noindent\textbf{AUTHOR DECLARATIONS}\\
\textbf{Conflict of Interest}\\
The authors have no conflicts to disclose.\\

\noindent\textbf{Author Contributions}\\
S.D. and C.Q. proposed the initial idea for the proposed architecture. S.D., Y.D. and C.Q. conceived the paper, performed data analysis, and discussed the results in the paper.\\

\noindent\textbf{Data Availability}\\
The data supporting the findings of this study is available either within the paper or from the corresponding author upon reasonable request. \\

\noindent\textbf{Code Availability}\\
A simulator is developed by Y.D. to generate the data points plotted in Figures 5 and 6. The code for this simulator is available  from the corresponding author upon reasonable request. 

\bibliography{SQGPU}

\end{document}